\documentclass{pasj01}

\RequirePackage{mathpazo}
\RequirePackage[T1]{fontenc}

\newcounter{author}
\def\authorcount#1#2{\refstepcounter{author}\label{#1}
                     \altaffiltext{\ref{#1}}{#2}}

\def\kojiguchiprep{N. Kojiguchi et al. in preparation}

\begin{document}
\SetRunningHead{T. Kato and N. Kojiguchi}{New Candidates for AM CVn Stars}

\Received{202X/XX/XX}
\Accepted{202X/XX/XX}

\title{New Candidates for AM Canum Venaticorum Stars among
      ASAS-SN Transients}

\author{Taichi~\textsc{Kato},\altaffilmark{\ref{affil:Kyoto}*}
        Naoto~\textsc{Kojiguchi}\altaffilmark{\ref{affil:Kyoto}}
}

\authorcount{affil:Kyoto}{
     Department of Astronomy, Kyoto University, Kyoto 606-8502, Japan}
\email{$^*$tkato@kusastro.kyoto-u.ac.jp}


\KeyWords{accretion, accretion disks
          --- stars: novae, cataclysmic variables
          --- stars: dwarf novae
          --- stars: individual (ASASSN-20eq, ASASSN-20gx, ASASSN-20jt,
              ASASSN-20ke, ASASSN-20la, ASASSN-20lr)
          --- surveys
         }

\maketitle

\begin{abstract}
We studied Zwicky Transient Facility (ZTF) light curves
of 34 dwarf nova candidates discovered by All-Sky Automated Survey
for Supernovae (ASAS-SN) between 2020 May 12 and September 9
and found 6 AM CVn-type candidates.
All objects showed short outbursts (post-superoutburst
rebrightenings) on the fading tail.  Two objects
(ASASSN-20eq, ASASSN-20la) showed double superoutbursts.
Three objects (ASASSN-20jt, ASASSN-20ke, and ASASSN-20lr)
showed short superoutbursts (5--6~d).
These features in the light curve can be
used in discriminating AM CVn-type candidates from
hydrogen-rich systems.  In contrast to hydrogen-rich systems,
some object did not show red color excess during the rebrightening
or fading tail phase.  We interpret that this is due
to the higher ionization temperature in helium disks.
Two objects had long (likely) supercycles:
ASASSN-20gx (8.5~yr) and ASASSN-20lr (7~yr).
We provide a scheme for identifying AM CVn-type candidates
based on the light curve characteristics.
\end{abstract}

\section{Introduction}\label{sec:intro}

   AM CVn stars are a class of cataclysmic variables (CVs)
containing a white dwarf (primary) and a mass-transferring
helium white dwarf (secondary).
[For a review of AM CVn stars, see e.g. \citet{sol10amcvnreview}].
In systems with mass-transfer rates ($M_{\rm dot}$)
the accretion disk around the primary become thermally stable 
and no outbursts are observed.  In systems with
lower $M_{\rm dot}$, the disk becomes thermally unstable
and dwarf nova (DN)-type outbursts occur
(\cite{tsu97amcvn}; \cite{sol10amcvnreview}; 
\cite{kot12amcvnoutburst}).
The mass-transfer in AM CVn stars is driven by
angular momentum loss due to the gravitational wave radiation
and $M_{\rm dot}$ is a strong function of the orbital period
($P_{\rm orb}$).  Systems with short $P_{\rm orb}$
(less than 22 min) have thermally stable disks and
those with longer $P_{\rm orb}$ have thermally unstable
disks (\cite{ram12amcvnLC}; \cite{kot12amcvnoutburst}).
It is not yet clear whether AM CVn stars with
extremely low $M_{\rm dot}$ have thermally unstable disks
and show outbursts, but the recent discovery of
an outbursting AM CVn star, ZTF20acyxwzf,
with $P_{\rm orb}$=0.0404(3)~d (\kojiguchiprep)
suggests that AM CVn stars even with the longest $P_{\rm orb}$
have thermally unstable disks.

   AM CVn stars have low mass-ratios ($q=M_2/M_1$).
In systems with low $q$, the disk becomes tidally unstable
due to the 3:1 resonance (\cite{whi88tidal};
\cite{hir90SHexcess}; \cite{lub91SHa}) and
the precessing eccentric disk whose eccentricity is
excited by the 3:1 resonance causes superhumps
and superoutbursts \citep{osa89suuma}.  In extremely
low-$q$ systems, the disk can even hold the radius of
the 2:1 resonance and this is believed to be responsible
for the WZ Sge-type phenomenon \citep{osa02wzsgehump},
which show infrequent large-amplitude superoutbursts
and often post-superoutburst rebrightenings \citep{kat15wzsge}.
Post-superoutburst rebrightenings in AM CVn stars
are relatively commonly seen (\cite{iso15asassn14ei};
\cite{duf21amcvn}).

   AM CVn stars have recently receiving special attention
and there have been a number of projects in search of
AM CVn stars.  \citet{and05SDSSamcvn} and \citet{rau10HeDN}
used spectra and \citet{car13SDSSamcvn} used colors in
the Sloan Digital Sky Survey (SDSS).  \citet{lev15amcvn} used
the Palomar Transient Factory (PTF) to detect outbursting
AM CVn stars.  High time-resolution observations also discovered
several AM CVn stars [e.g. \citet{bur19j1539} and \citet{bur20amcvn}
using the Zwicky Transient Facility (ZTF) data].
A number of outbursting AM CVn stars have been identified
by time-series observations to search for superhumps
(e.g. \cite{kat15asassn14cc}; \cite{iso19nsv1440};
\cite{iso15asassn14ei}).
Most recently, \citet{vanroe21amcvn} selected
ZTF transients by colors and detected several new
AM CVn stars.

   In this paper, we present new candidate AM CVn stars
from recently detected potential dwarf novae (DNe) by
the All-Sky Automated Survey for Supernovae (ASAS-SN)
(\cite{ASASSN}; \cite{koc17ASASSNLC}).\footnote{
   The list of ASAS-SN Transients is available at
$<$http://www.astronomy.ohio-state.edu/$^\sim$assassin/transients.html$>$.
} using the public light curves of the ZTF survey\footnote{
   The ZTF data can be obtained from IRSA
$<$https://irsa.ipac.caltech.edu/Missions/ztf.html$>$
using the interface
$<$https://irsa.ipac.caltech.edu/docs/\\
program\_interface/ztf\_api.html$>$
or using a wrapper of the above IRSA API
$<$https://github.com/MickaelRigault/ztfquery$>$.
}.
We supplemented the data using the Asteroid Terrestrial-impact
Last Alert System (ATLAS) Forced Photometry
\citep{ATLAS}\footnote{
  The ATLAS Forced Photometry is available at
  $<$https://fallingstar-data.com/forcedphot/$>$.
}.
The list of object is shown in table \ref{tab:obj}.
The coordinates and variability range are taken from
AAVSO VSX.\footnote{
  $<$https://www.aavso.org/vsx/$>$.
}).  The parallaxes and quiescent magnitudes are taken
from Gaia EDR3 \citep{GaiaEDR3}

\begin{table*}
\caption{List of objects}\label{tab:obj}
\begin{center}
\begin{tabular}{ccccccc}
\hline
Object & Right Ascention & Declination & Variability & Parallax (mas) & Gaia & Gaia \\
       & (J2000.0)       & (J2000.0)   & Range       &                & BP   & RP \\
\hline
ASASSN-20eq & \timeform{17h 35m 00.45s} & \timeform{+25D 36' 54.8''} & 15.4--21.5g & -- & -- & -- \\
ASASSN-20gx & \timeform{23h 49m 30.14s} & \timeform{+22D 01' 29.6''} & 14.8--20.3g & 1.40(52) & 20.40(10) & 19.91(13) \\
ASASSN-20jt & \timeform{23h 02m 36.07s} & \timeform{+53D 28' 22.6''} & 17.0--22.1g & -- & 20.12(10) & 19.36(11) \\
ASASSN-20ke & \timeform{18h 42m 17.83s} & \timeform{+16D 55' 01.3''} & 16.1--21.0g & 0.95(78) & 20.89(10) & 20.39(21) \\
ASASSN-20la & \timeform{01h 38m 51.95s} & \timeform{+46D 34' 48.9''} & 16.1--21.5g & -- & -- & -- \\
ASASSN-20lr & \timeform{04h 22m 20.06s} & \timeform{+50D 07' 13.0''} & 14.6--20.0g & 2.19(40) & 19.83(3) & 19.38(5) \\
\hline
\end{tabular}
\end{center}
\end{table*}

\section{Individual Objects}\label{sec:individual}

\subsection{ASASSN-20eq}

   This object was detected at $g$=15.6 on 2020 May 12.
We noticed that this object showed multiple rebrightenings
on a fading tail from ZTF observations.  The quiescent color
in SDSS is unusual in that the object had a strong
ultraviolet excess of $u-g$=$-$0.22 (vsnet-alert 25852).\footnote{
  The vsnet-alert messages can be seen at
  $<$http://ooruri.kusastro.kyoto-u.ac.jp/pipermail/vsnet-alert/$>$.
}
By supplying ATLAS and ASAS-SN observations, D. Denisenko
(vsnet-alert 25859) and P. Schmeer (vsnet-alert 25861)
identified the initial superoutburst.

\begin{figure*}
  \begin{center}
    \FigureFile(150mm,110mm){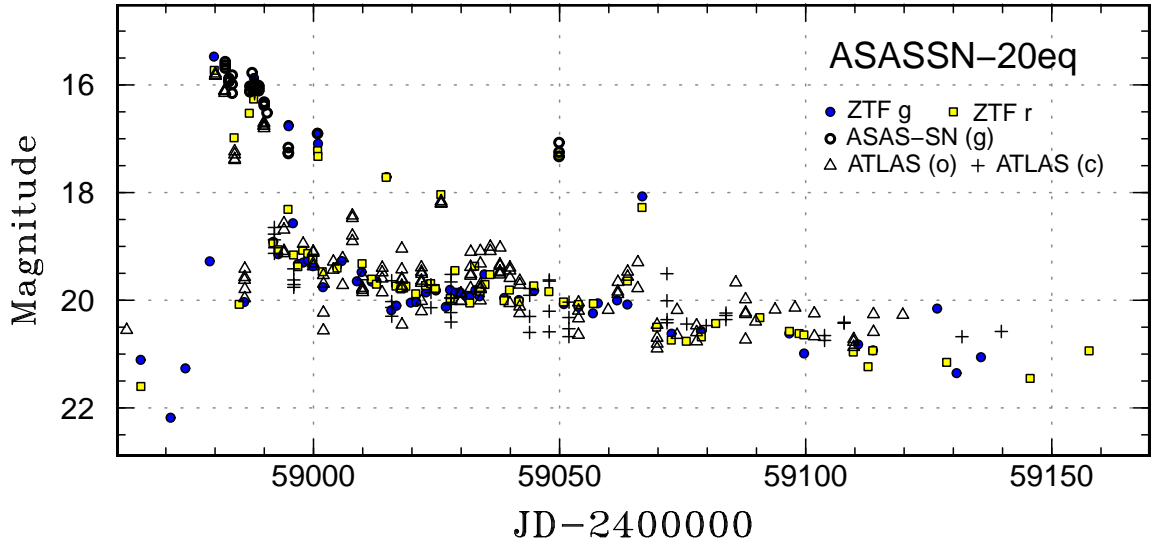}
  \end{center}
  \caption{Light curve of ASASSN-20eq.
  The filled squares and circles represent ZTF $r$ and $g$
  observations, respectively.
  Open circles, triangles and ``+'' signs represent
  ASAS-SN, ATLAS o and ATLAS c observations, respectively.
  }
  \label{fig:asn20eqlc}
\end{figure*}

   The combined light curve (figure \ref{fig:asn20eqlc})
shows two superoutbursts (JD 2458979--2456984 and
JD 2458986--2458991) separated by fading.
Six post-superoutburst rebrightenings were detected
in the available data.

   The overall light curve is very similar to that of
``double superoutburst'' of an AM CVn star NSV 1440
\citep{iso19nsv1440}.  Very rapid fading (more than
2 mag d$^{-1}$) of rebrightenings is also characteristic
of AM CVn-type outbursts [there is Bailey relation
for hydrogen-rich dwarf novae: the decline rate
$T_{\rm decay}$ is proportional to $P_{\rm orb}^{0.79}$
(\cite{war87CVabsmag}; \cite{war95book}) and it has been
confirmed to apply to AM CVn stars \citep{pat97crboo}].
Based on these characteristics of the light curve and
a strong ultraviolet excess, we identified this object
to be an AM CVn star.  The initial superoutburst was
most likely characterized by the 2:1 resonance
(\cite{iso19nsv1440}; \cite{Pdot5}) and the second one
almost certainly showed ordinary superhumps.
Based on the similarity of the light curve with that of
NSV 1440, the $P_{\rm orb}$ of ASASSN-20eq is expected
to be around 0.025~d.

\subsection{ASASSN-20gx}

\begin{figure*}
  \begin{center}
    \FigureFile(150mm,110mm){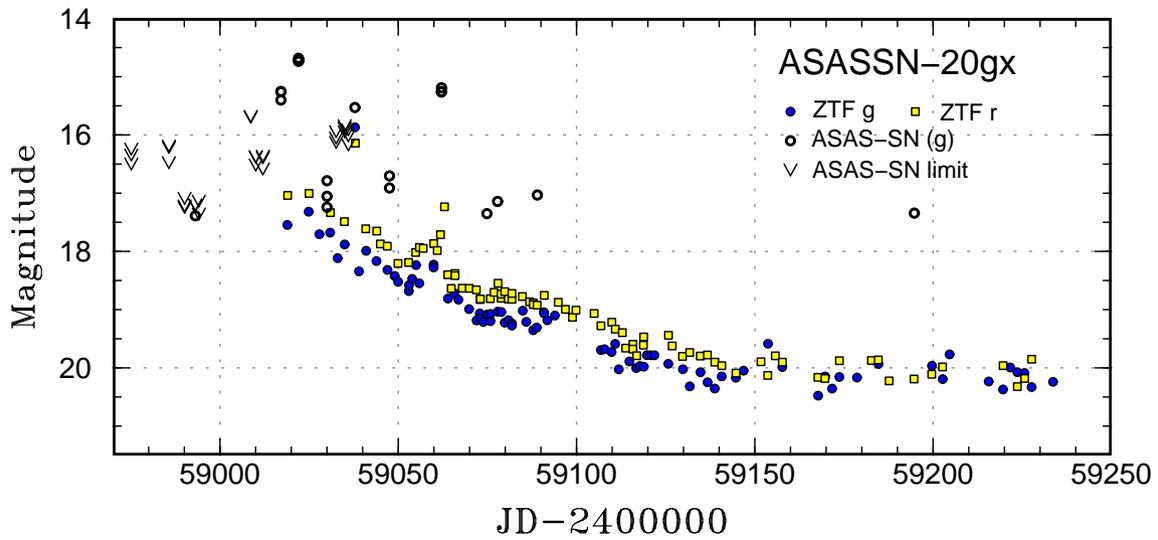}
  \end{center}
  \caption{Light curve of ASASSN-20gx.
  The filled squares and circles represent ZTF $r$ and $g$
  observations, respectively.
  Open circles and ``v'' marks represent ASAS-SN observations
  and upper limits, respectively.
  }
  \label{fig:asn20gxlc}
\end{figure*}

   This object was detected at $g$=15.4 on 2020 June 16
and further brightened to $g$=14.8 on 2020 June 21.
We noticed multiple rebrightenings on a fading tail
from ZTF observations as in ASASSN-20eq (vsnet-alert 25853).
The light curve (figure \ref{fig:asn20gxlc})
suggests that ASAS-SN observations missed the initial
superoutburst during observational or seasonal gaps
(maximum of a 12~d gap and the observations started
just after the seasonal gap).

   There were at least five rebrightenings
(assuming that there was an unrecorded superoutburst).
These rebrightenings showed rapid fading (2 mag d$^{-1}$).
Combined with the blue color in quiescence ($u-g$=$+$0.15
in SDSS), we consider that this object is also
an AM CVn star.
Although we cannot completely exclude
the complete absence of the initial superoutburst from
the available observations, this possibility appears
to be low considering the similarity of the light curve
of the fading tail with those of other well-observed
superoutbursts of AM CVn stars.

   D. Denisenko reported another past outburst in
December 2011--January 2012 (vsnet-alert 25860) in
Catalina Sky Survey Data (CRTS, \cite{CRTS}).
The Panoramic Survey Telescope and Rapid Response
System (Pan-STARRS1, \cite{PS1})\footnote{
  $<$https://panstarrs.stsci.edu/$>$.
} also detected this outburst
and CRTS probably recorded the fading tail.
If there was not outburst between 2012 and 2020,
the supercycle is probably around 8.5~yr.

\subsection{ASASSN-20jt}

\begin{figure*}
  \begin{center}
    \FigureFile(150mm,110mm){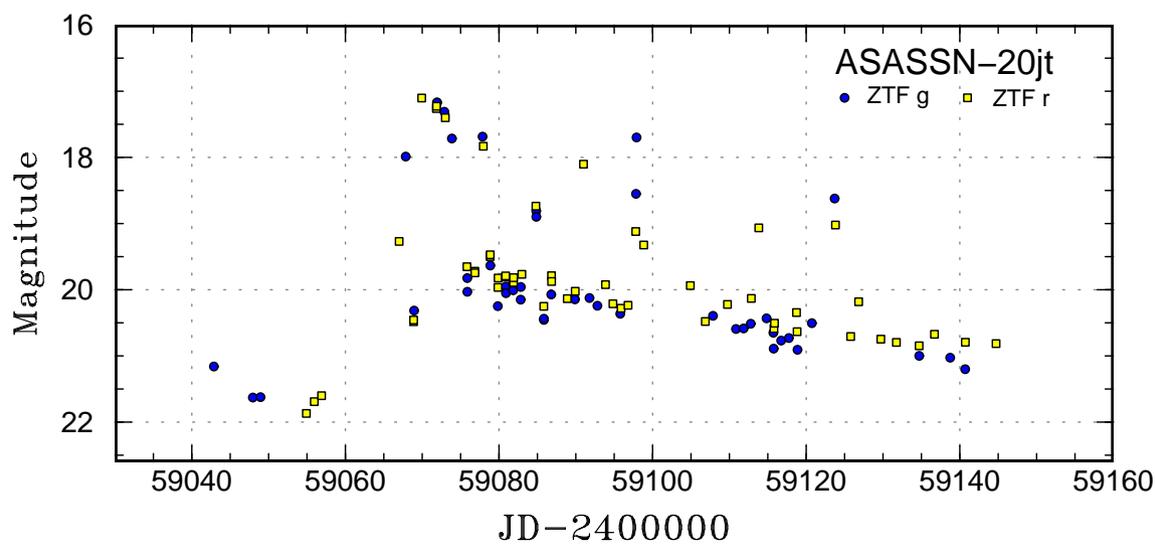}
  \end{center}
  \caption{Light curve of ASASSN-20jt.
  The filled squares and circles represent ZTF $r$ and $g$
  observations, respectively.
  }
  \label{fig:asn20jtlc}
\end{figure*}

   This object was detected at $g$=17.0 on 2020 August 7.
The light curve based on the ZTF data
(figure \ref{fig:asn20jtlc}) indicates brightening
from $r$=19.27 on 2020 August 5 (JD 2459067) to $g$=18.00 on
2020 August 6, followed by a dip at $g$=20.32 on
2020 August 7 (despite the ASAS-SN transient detection,
no positive observation was available from
the ASAS-SN Sky Patrol).
After this, a long outburst lasting 
at least 4~d was recorded.  There were six rebrightenings
on a fading tail.  The initial short outburst was likely
a precursor and the long outburst was likely
a superoutburst.  Based the short duration of
the main superoutburst and rapid fading (up to 1.7 mag d$^{-1}$),
we identified this object as a likely AM CVn star.

   There was a similar, but less observed, outburst in
2018 October--December in the ZTF data.  The observations
only recorded the phase of fading tail
and three rebrightenings were detected on it.
The supercycle of this object is estimated to be
$\sim$670~d.

\subsection{ASASSN-20ke}

\begin{figure*}
  \begin{center}
    \FigureFile(150mm,110mm){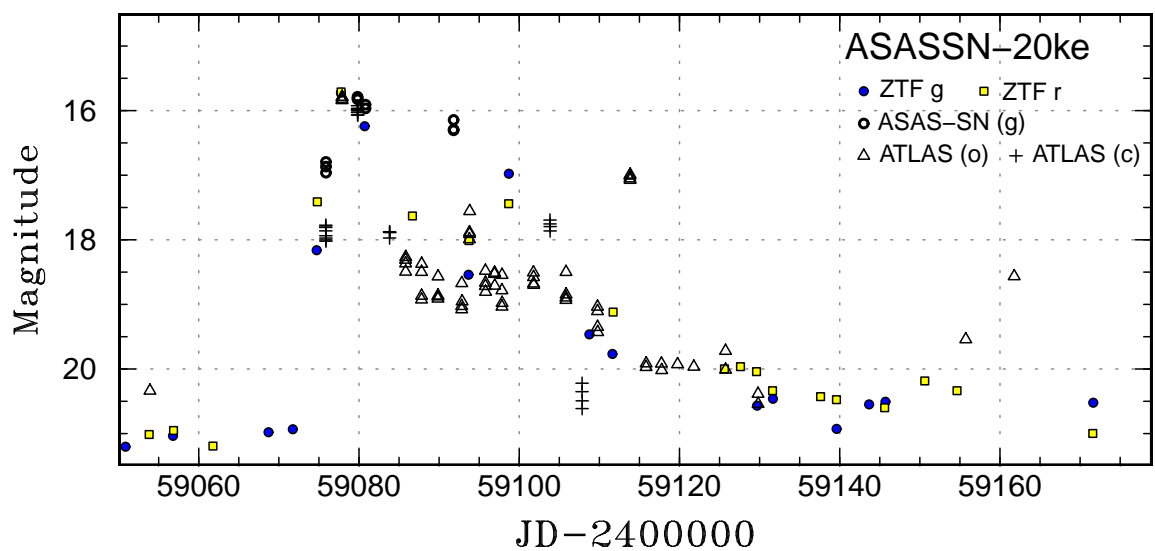}
  \end{center}
  \caption{Light curve of ASASSN-20ke.
  The filled squares and circles represent ZTF $r$ and $g$
  observations, respectively.
  Open circles, triangles and ``+'' signs represent
  ASAS-SN, ATLAS o and ATLAS c observations, respectively.
  }
  \label{fig:asn20kelc}
\end{figure*}

   This object was detected at $g$=16.2 on 2020 August 18.
The light curve based on the ZTF and ASAS-SN data
(figure \ref{fig:asn20kelc}) indicates the initial
long outburst lasting 6~d.
There was at least three confirmed post-superoutburst
rebrightenings.  Although there were several more ASAS-SN
detections around $g$=17.0,
they were spurious detections near the detection limit
which were confirmed by comparison with ATLAS data.
These detections were not plotted on the figure.
This object is most likely classified as an AM CVn star
based on the short duration of the initial superoutburst
and rapid fading (more than 2 mag d$^{-1}$) of
rebrightenings.

   There was another outburst in 2019 July-August.
This outburst was only detected by ZTF and ATLAS
and rebrightenings on a long-lasting (at least 70~d)
fading tail were recorded.
The initial part of this outburst was not recorded
due to the long observational gap.  The supercycle
of this object is estimated to be $\sim$410~d.

\subsection{ASASSN-20la}

\begin{figure*}
  \begin{center}
    \FigureFile(150mm,110mm){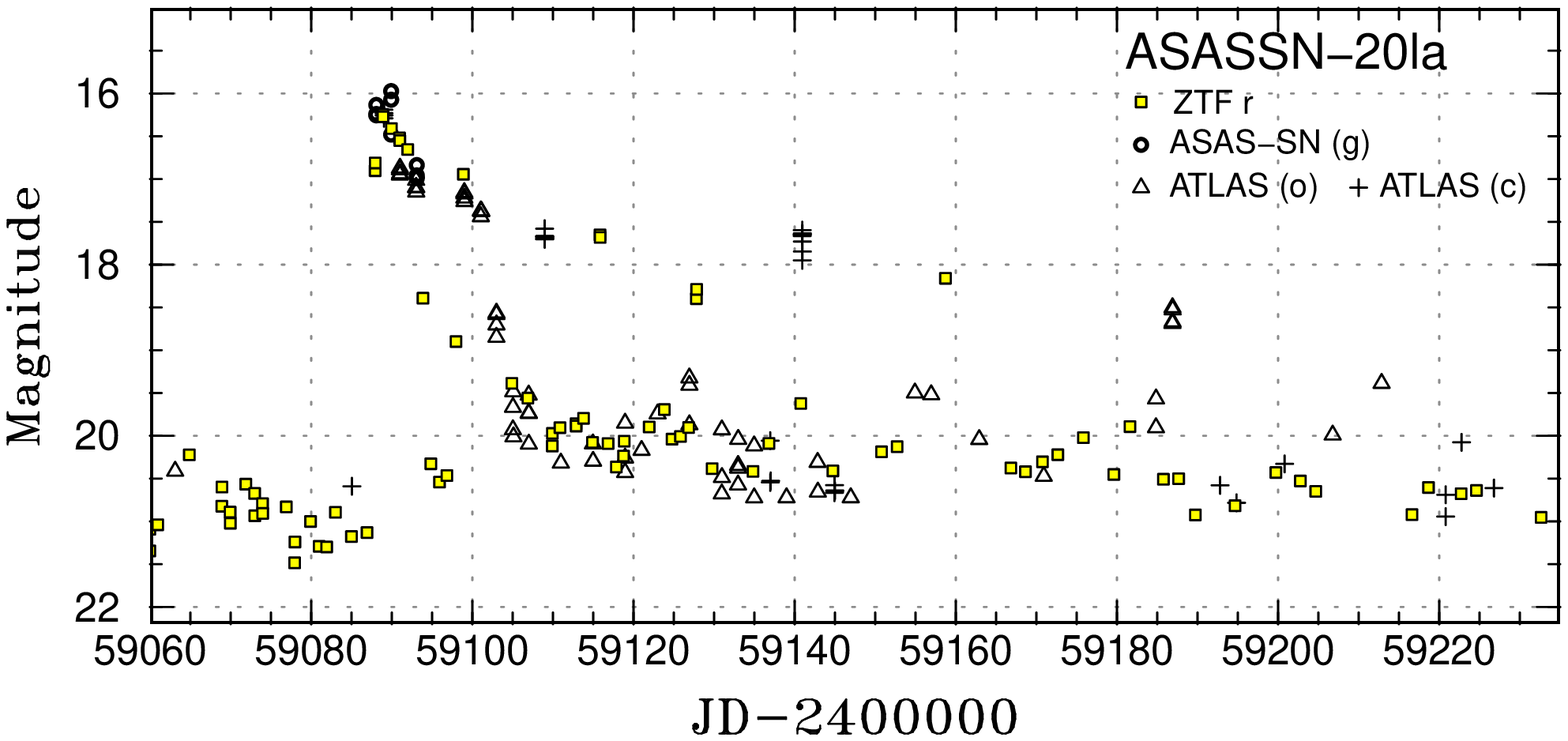}
  \end{center}
  \caption{Light curve of ASASSN-20la.
  The filled squares represent ZTF $r$ observations.
  Open circles, triangles and ``+'' signs represent
  ASAS-SN, ATLAS o and ATLAS c observations, respectively.
  }
  \label{fig:asn20lalc}
\end{figure*}

   This object was detected at $g$=16.1 on 2020 August 28.
The light curve based on the ZTF, ASAS-SN and ATLAS
Forced Photometry data (figure \ref{fig:asn20lalc})
indicates the initial
superoutburst lasting 6~d (JD 2459088--2459064)
followed by a dip, and the possible second superoutburst
(JD 2459098--2459101).  Six post-superoutburst
rebrightenings were detected on the fading tail.
The shortness of the initial superoutburst is
incompatible with a hydrogen-rich DN.  Likely
double superoutburst and the rapid fading rate
(more than 2 mag d$^{-1}$) of rebrightenings
also support the AM CVn-type classification.
No previous outburst was detected in ASAS-SN
(since 2013 November) and ZTF (since 2018 June).
Even considering the seasonal observational gaps,
the lack of previous signature of a fading tail
suggests that the supercycle is longer than 900~d.

\subsection{ASASSN-20lr}

\begin{figure*}
  \begin{center}
    \FigureFile(150mm,110mm){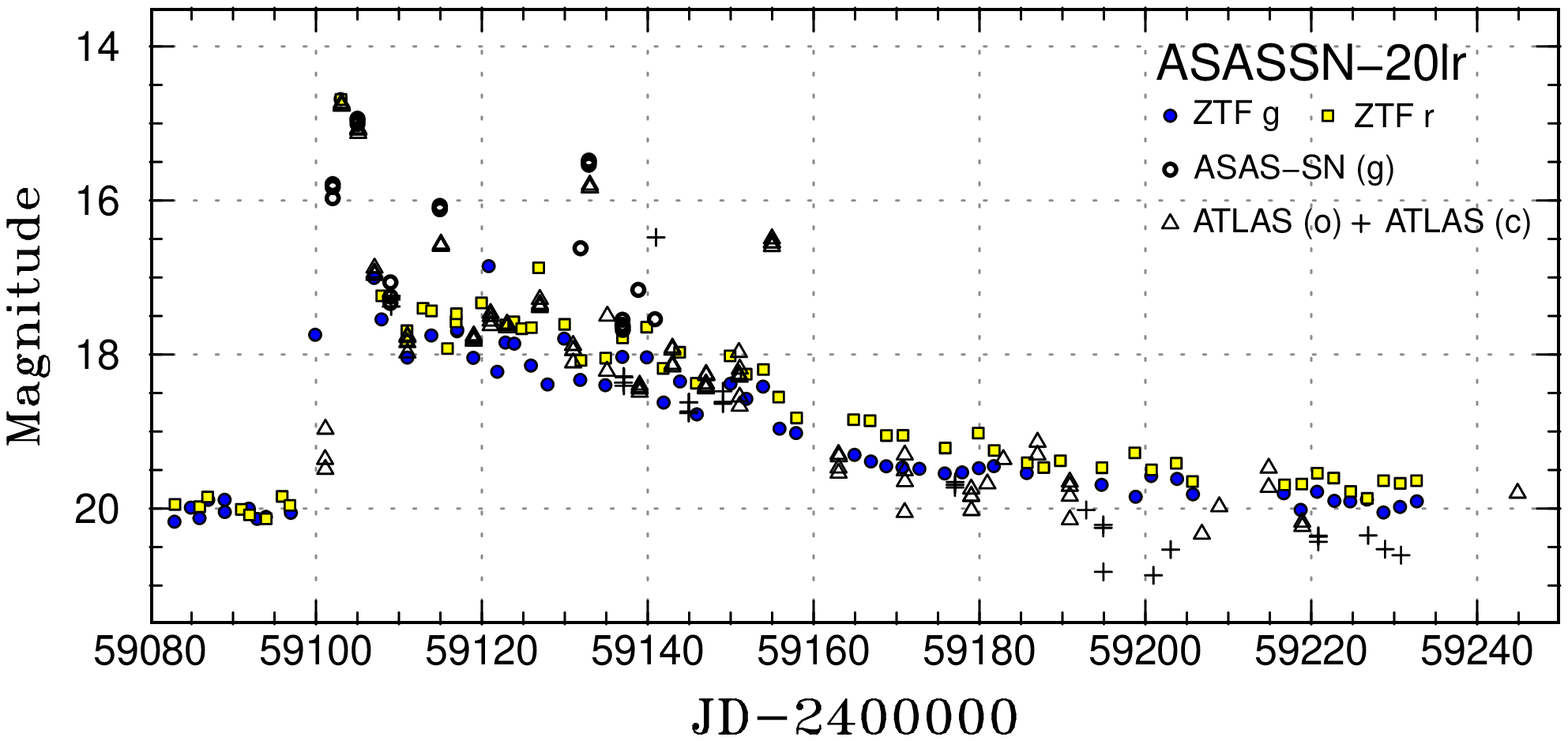}
  \end{center}
  \caption{Light curve of ASASSN-20lr.
  The filled squares and circles represent ZTF $r$ and $g$
  observations, respectively.
  Open circles, triangles and ``+'' signs represent
  ASAS-SN, ATLAS o and ATLAS c observations, respectively.
  }
  \label{fig:asn20lrlc}
\end{figure*}

   This object was detected at $g$=15.9 on 2020 September 9.
The light curve based on the ZTF and ASAS-SN data
(figure \ref{fig:asn20lrlc}) indicates the initial
superoutburst lasting 5~d (JD 2459100--2459105).
There were at least three post-superoutburst rebrightenings.
As in ASASSN-20la, the short duration
of the initial superoutburst and the rapid
fading rate (more than 2 mag d$^{-1}$) of rebrightenings
support the AM CVn-type classification.
Pan-STARRS1 data recorded a fading tail in 2016
and the supercycle of this object is estimated to be
$\sim$7~yr.

\section{Discussion}\label{sec:discuss}

\subsection{Number statistics}

   A total of six AM CVn candidates discovered within
four months is amazingly high in number.
They comprised 18\% of 34 newly discovered ASAS-SN DN
candidates during the same interval having ZTF light curves
which had a good temporal coverage and quality allowing
type classification.  The total number of ASAS-SN DN candidates
during the same interval was 111.
The ratio of 18\% appears too high to reflect
the population statistics among DNe, and this high
number may have simply been a result of random
fluctuation.  The estimated parent fraction of AM CVn candidates
has a 95 percent confidence interval of [0.068, 0.345].
In fact, in our previous survey of SU UMa-type DNe
\citep{Pdot7}, we found that 8\% of objects showing dwarf
nova-type outbursts were AM CVn-type objects (8 out of
105 outbursts; all of them were confirmed either by
spectroscopy or by the detection of superhumps or eclipses).
The 95 percent confidence interval of the parent fraction
of AM CVn stars in this sample was [0.033, 0.145].
This interval overlaps with the estimate in the present
study.
Combined with the sequence of recent discoveries
of AM CVn stars or candidates
[ASASSN-21au = ZTF20acyxwzf \citep{iso21asassn21auatel},
ASASSN-21eo (vsnet-alert 25635) and
ASASSN-21hc (vsnet-alert 25849, 25868)] among
ASAS-SN DN candidates, the recent statistics of outbursting
objects may suggest a signature of a larger fraction of
AM CVn stars among CVs than had previously been thought
(e.g. about 1\% in RKcat Edition 7.21 \cite{RKCat}).

   There may have also been selection biases,
such as the past detection scheme in ASAS-SN more easily
detected hydrogen-rich DNe than helium ones
(e.g. the short duration of superoutbursts in helium DNe
would make detections more difficult in low-cadence
surveys).  Such a potential bias needs to be examined
in more detail in discussing the fraction of AM CVn stars
among DNe.

\subsection{Post-superoutburst rebrightenings and fading tail}

   In hydrogen-rich CVs, multiple post-superoutburst rebrightenings
are associated with WZ Sge-type DNe \citep{kat15wzsge}, but
no exclusively [e.g. V1006 Cyg, \citet{kat16v1006cyg}; ASASSN-14ho
\citet{kat20asassn14ho}].  Although the cause of these
rebrightenings is still poorly understood, the infrared or red
excess observed during the rebrightening phase or
during the fading tail in hydrogen-rich systems
(\cite{uem08j1021}; \cite{mat09v455and};
\cite{cho12j2138}; \cite{nak13j0120}; \cite{gol14j1915proc};
\cite{iso15ezlyn}) is usually considered to arise from
an optically thin region that is located outside the optically
thick disk, and this matter in the outer disk would serve
as a mass reservoir (\cite{kat98super}; \cite{kat04egcnc})
to enable rebrightenings.

   Among the AM CVn-type candidate we studied, ASASSN-20eq
and ASASSN-20jt did not show significant red colors
(ZTF $g-r$) during the rebrightening/fading tail phase
(we refer to the colors in comparison with those around
the outburst peak or in quiescence).
This apparently makes clear contrast to hydrogen-rich systems.
This may be a result of higher ionization temperature
of helium compared to hydrogen, and optically thin region
can still emit bluer light compared to hydrogen-rich systems.
These instances suggest that the lack of red excess during
the rebrightening/fading tail phase can be used for
identifying AM CVn stars.

   Two objects (ASASSN-20gx and ASASSN-20lr) showed
a some degree of red color during the same phase.
These instances suggest that the outer part of the disk
can become cool enough to emit red light even in
helium systems.

\subsection{Implication on transient selection}

   As introduced in section \ref{sec:intro} AM CVn stars
have been receiving much attention in recent years.
Selections of AM CVn stars from other CVs, however,
have been a challenge in most cases.  Since the fraction
of AM CVn stars among CVs is low, there always arises
a serious problem of detecting a small number of objects
among the far numerous non-AM CVn background.  This usually
causes a large number of false positives (undesired hydrogen-rich
systems) if the criterion is loose.  With a more stringent
criterion, many false negatives (many AM CVn stars classified
as ordinary CVs) occur.  The small number of AM CVn samples
also would provide a difficult condition in machine learning.

   Using one of the criteria (BP$-$RP $<$ 0.6) in \citet{vanroe21amcvn},
ASASSN-20jt (BP$-$RP=$+$0.76) becomes a false negative
among the three objects with known Gaia colors.
Using their criterion of high priority
candidates ($-$0.6 $<$ BP$-$RP $<$ 0.3) all three objects
with Gaia colors in our sample are not considered as
high priority.
This indicates the limitation in choosing candidates
by colors only (particularly in the presence of
high number of ``background'' hydrogen-rich objects).

   We propose to use the structure of the light curve
as a better selection tool of AM CVn-type candidates.
We also added some additional features useful for
identifying AM CVn-type outbursts.

They are:
\begin{enumerate}

\item Rapid fading (more than 1.5 mag d$^{-1}$ during
any part of the light curve).

\item Short duration (usually 5--6~d) of the superoutburst.
In hydrogen-rich systems, superoutbursts usually
last more than 10~d.

\item Double superoutburst.  Double superoutbursts
are rare in hydrogen-rich systems and initial superoutburst
of the double superoutburst last longer than 10~d
(see e.g. \cite{kat13j1222}).  Rapid fading (item 1)
after $\sim$5~d-long outburst (item 2) is a strong
sign of an AM CVn system and observations to watch for
the second superoutburst and emergence of superhumps
are very desirable.

\item Long fading tail lasting 100--200~d despite
the lack of long (usually more than 20~d in hydrogen-rich
systems) superoutburst.
The lack of red excess in this stage would also be
a signature of an AM CVn system.

\item Outburst amplitudes of long outburst (4--6~mag)
smaller than hydrogen-rich WZ Sge stars (6--8~mag or
even more).  This reflects the small disk size 
in AM CVn-type objects.  Potential confusions with outbursts in
hydrogen-rich systems with lower amplitudes (such as
SU UMa stars or SS Cyg stars) could be
avoided by confirming the absence of past outbursts.

\item In the same sense, a faint absolute magnitude
(significantly fainter than $+$4) of a long outburst
can be a signature of an AM CVn-type superoutburst,
if the parallax is known.  For example, the maximum
absolute of ASASSN-20lr is $+$6.3(4).

\end{enumerate}

   In summary, item 1 is probably most useful in practice.
If the number of observations is sufficient, item 2
is also very helpful.  If the light curve is known
long after the event, items 3 and 4 will be helpful.
Items 5 and 6 will be helpful if the quiescent
counterpart can be identified or Gaia parallax
is available.

   Incorporation of these features in automated recognizing
system will certainly increase the success rate
in follow-up spectroscopic observations.

   Upon request by the referee Michael Coughlin, we provide
a toy R code to implement the items 1 and 2.  Using the actual
ZTF $r$ data, this code correctly recognized ASASSN-20eq,
ASASSN-20jt, ASASSN-20ke (for the 2019 outburst) and ASASSN-20la
as AM CVn-type superoutbursts while the data for the hydrogen-rich
WZ Sge star AL Com did not pass this test.  The reason why
ASASSN-20gx did not pass the test was due to
the observational gaps in the ZTF data, causing an apparent
fading rate smaller than 1.5 mag d$^{-1}$
(if we loosen the criterion to 1.2 mag d$^{-1}$, this object
is recognized as an AM CVn star).  The reason why ASASSN-20lr
did not pass the test was the lack of observations immediately
after the peak.  The second observation by the ZTF was 4~d
after the peak and it was impossible to measure the duration
of the initial outburst only by the ZTF data. 
We hope others would benefit from this toy code
and perhaps have ideas to turn it into a better filter.

\section*{Acknowledgments}

This work was supported by JSPS KAKENHI Grant Number 21K03616.

We are also grateful to the ASAS-SN, CRTS and Pan-STARRS
teams for making the database available to the public.

Based on observations obtained with the Samuel Oschin 48-inch
Telescope at the Palomar Observatory as part of
the Zwicky Transient Facility project. ZTF is supported by
the National Science Foundation under Grant No. AST-1440341
and a collaboration including Caltech, IPAC, 
the Weizmann Institute for Science, the Oskar Klein Center
at Stockholm University, the University of Maryland,
the University of Washington, Deutsches Elektronen-Synchrotron
and Humboldt University, Los Alamos National Laboratories, 
the TANGO Consortium of Taiwan, the University of 
Wisconsin at Milwaukee, and Lawrence Berkeley National Laboratories.
Operations are conducted by COO, IPAC, and UW.

The ztfquery code was funded by the European Research Council
(ERC) under the European Union's Horizon 2020 research and 
innovation programme (grant agreement n$^{\circ}$759194
-- USNAC, PI: Rigault).

This work has made use of data from the Asteroid Terrestrial-impact
Last Alert System (ATLAS) project. The Asteroid Terrestrial-impact 
Last Alert System (ATLAS) project is primarily funded to search for 
near earth asteroids through NASA grants NN12AR55G, 80NSSC18K0284, 
and 80NSSC18K1575; byproducts of the NEO search include images and 
catalogs from the survey area. This work was partially funded by 
Kepler/K2 grant J1944/80NSSC19K0112 and HST GO-15889, and STFC grants 
ST/T000198/1 and ST/S006109/1. The ATLAS science products have been 
made possible through the contributions of the University of 
Hawaii Institute for Astronomy, the Queen's University Belfast,
the Space Telescope Science Institute, the South African Astronomical 
Observatory, and The Millennium Institute of Astrophysics (MAS), Chile.

This research has made use of the International Variable Star Index 
(VSX) database, operated at AAVSO, Cambridge, Massachusetts, USA.

\section*{Supporting information}

   The toy R code to detect AM CVn-type outbursts can be found
in the online version of this article. \\
Supplementary data is available at PASJ Journal online.

\end{document}